\documentclass[aps,preprint]{revtex4}
\usepackage{amsfonts}
\usepackage{amssymb}

\begin{document}

\title{Theory of the ferroelectric phase in organic conductors:\\
optics and physics of solitons.}
\author{S. Brazovskii}
\affiliation{
Laboratoire de Physique Th\'{e}orique et des Mod\`{e}les Statistiques, CNRS,
B\^{a}t.100, Universit\'{e} Paris-Sud, 91406 Orsay cedex, France \\
\textrm{and}\\
L.D.Landau Institute, Moscow, Russia\\
\textrm{e-mail: brazov@ipno.in2p3.fr}
}
\published[Written for
proceedings of the ECRYS-2002 \cite{ecrys-02}; the updated and
extended version of \cite{brazov-ecrys-02}]
\bigskip

\begin{abstract}
Recently the ferroelectric anomaly (Nad, Monceau, et al) followed
by the charge disproportionation (Brown, et al) have been
discovered in $(TMTTF)_{2}X$ compounds. The corresponding theory
of the combined Mott-Hubbard state describes both effects by
interference of the build-in nonequivalence of bonds and the
spontaneous one of sites. The state gives rise to three types of
solitons: $\pi -$ solitons (holons) are observed via the
activation energy $\Delta $ in the conductivity $G$; noninteger
$\alpha - $ solitons (the FE domain walls) provide the frequency
dispersion of the ferroelectric response; combined spin-charge
solitons determine $G(T)$ below subsequent structural transitions
of the tetramerisation. The photoconductivity gap $2\Delta $ is
determined by creations of soliton - antisoliton pairs. The
optical edge lies well below, given by the collective
ferroelectric mode which coexists with the combined
electron-phonon resonance and the phonon antiresonance. The charge
disproportionation and the ferroelectricity can exist hiddenly
even in the $Se$ subfamily giving rise to the unexplained yet low
frequency optical peak, the enhanced pseudogap and traces of
phonons activation.
\bigskip
\bigskip
\end{abstract}

\maketitle


\section{Combined Mott-Hubbard state, charge disproportionation and
ferroelectricity.}

The family of quasi one-dimensional organic superconductors
(Bechgard - Fabre salts $(TMTSF)_{2}X$ and $(TMTTF)_{2}X$)
demonstrates, at low temperatures, transitions to almost all known
electronic phases \cite{jerome}. At higher $T_{ao}$, usually there
is also a set of weak structural transitions of the anion
orderings (AOs) which are slight arrangements of chains of
counterions $X$ \cite{anions}. At even higher $T\approx T_{0}$,
also other structureless transitions \cite{lawersanne} were
observed sometimes (in the TMTTF subfamily), but they could not be
explained and later were left unattended with dramatic
consequences for the whole field. Recently their mysterious nature
has been elucidated by discoveries of the huge anomaly in the
dielectric susceptibility $\varepsilon $ \cite{nad,prl} and of the
charge disproportionation (CD) seen by the NMR \cite{brown}. The
new displacive instability and the usual orientational AOs seem to
be independent, as proved by finding their sequence in the
$(TMTTF)_{2}ReO_{4}$ \cite{nad}. The phase transition was
interpreted \cite{prl} as the least expected one: to the
Ferroelectric (FE) state, which is proved by the clear-cut fitting
of the anomaly in $\varepsilon (T)$ to the Curie law. The FE
transition is followed by a fast formation or a steep increase of
the conductivity gap $\Delta $ but with no appearance of the spin
gap. Hence we deal with a surprising FE version of the
Mott-Hubbard state which usually is associated rather with
magnetic orderings. The ferroelectricity was observed also in
dielectrical mixed-stacks organic compounds showing the
neutral-ionic transition (see \cite{rennes} for a short review).
Active experimental studies have been carried on during last two
years, particularly by methods of the NMR and the ESR
\cite{fujijama-02-03,nakamura-03}. (In references, we limit
ourselves to studies relevant to the FE transition, leaving aside
other cases of the CD.)

The FE transition in $(TMTTF)_{2}X$ is a very particular, bright
manifestation of a more general phenomenon of the CD, which
already has been predicted in \cite{seo-97} and now becomes
recognized as a common feature of various organic and some other
conductors, see \cite{fukuyama}. The phenomenon unifies a variety
of different concepts and observations, sometimes in quite unusual
aspects or conjunctions. Among them are the ferroelectricity of
good conductors, the instability towards the Mott-Hubbard state,
the Wigner crystallization \cite{kanoda} in a dense electronic
system, the ordered $4k_{F}$ density wave \cite{4K-F}, a richness
of physics of solitons, the interplay of structural and electronic
phases \cite{anions,physique+jetp}.

Already within the nonperturbed crystal structure the anions can
provoke the dielectrization. In $(TMTCF)_2X$ they dimerize
intermolecular distances, hence changing counting of the mean
electronic occupation from $1/2$ per molecule to $1$ per their
dimer. It originates \cite{zagreb} the small Umklapp scattering
$U_{b}$ which opens (according to Dzyaloshinskii \& Larkin, Luther
\& Emery) the route to the Mott-Hubbard insulator. While the bonds
are always dimerized, the molecules stay equivalent above $T_{0}$
which last symmetry is lifted by the CD. At $T<T_{0}$ the site
inequivalence adds more to $\Delta$ which is formed now by joint
effects of alternations of bonds and sites (remind also
\cite{matv} and \cite{0.5}). This change shows up as a kink in the
conductivity $G(T)$ at $T_{0}$ which turns down to higher $\Delta$
saturating at low $T$. The steepness of $G$ just below $T_{0} $
reflects the growth of the CD contribution to $\Delta$ which must
be correlated with $\varepsilon ^{-1}\sim T_{0}-T$. None of these
two types of dimerization change the unit cell of the zigzag stack
which basically contains two molecules, hence $q_{\parallel}=0$
($\mathbf{q}=(q_{\parallel},q_{\perp})$ is the CD wave vector).
But their sequence lifts the mirror and then the inversion
symmetries which must lead to the on-stack electric polarization.

By a good fortune, the 3D pattern of the CD appears in two,
anti-FE and FE, forms: \newline i) antiphase between stacks (found
only for $X=SCN$), here $q_{\perp }\neq 0$ which allows for its
structural identification \cite{anions};  \newline ii) inphase,
$\mathbf{q}=0$ hence the structureless character, which is the
macroscopic FE typically observed today \cite{nad}. \newline Both
types are the same paramagnetic insulators (the MI phase of
\cite{physique+jetp}); also their CD shows up similarly in the NMR
splitting \cite{brown}.

While the earliest theoretical approach \cite{seo-97} applies well
to a common situation \cite{fukuyama}, here the pronounced $1D$
electronic regime calls for a special treatment \cite{prl} which
also must be well suited to describe the FE properties. It is done
in terms of electronic phases $\varphi $ and $\theta $ (defined as
for the CDW order parameter $\sim \exp (i\varphi )\cos \theta $)
such that $\varphi ^{\prime }/\pi $ and $\theta ^{\prime }/\pi $
count local concentrations of the charge and the spin (see e.g.
\cite{reidel-fukuyama,reidel-maki} in \cite{reidel}). Beyond the
energies of charge and spin polarizations
\begin{equation}
\sim \hbar v_{F}(\varphi ^{\prime })^{2}\;\mathrm{and}\;\sim \hbar
v_{F}(\theta ^{\prime })^{2},  \label{phases}
\end{equation}
there are also the commensurability energies originated by site and bond
dimerizations (proportional to Umklapp amplitudes $U_{s}$ and $U_{b}$). At
presence of both of them, we arrive at the Hamiltonian for the combined
Mott-Hubbard state \cite{prl}
\[
H_{U}=-U_{s}\cos 2\varphi -U_{b}\sin 2\varphi =-U\cos (2\varphi -2\alpha
)~,~U=\sqrt{U_{s}^{2}+U_{b}^{2}}~,~\tan 2\alpha =U_{b}/U_{s}
\]
Quantum fluctuations renormalize $U$ down to $U^{\ast }(\neq 0$ at
$\gamma <1 $) which determines the gap $\Delta \sim
U^{1/(2-2\gamma )}$ as $U^{\ast }\sim \Delta ^{2}/\hbar v_{F}$.
The appearance of $U_{s}$ is regulated by one parameter $\gamma $
(the same as $\gamma _{\rho }$ of \cite{physique+jetp} or $K_{\rho
}$ of our days) which depends on electronic interactions. The
spontaneous CD $U_{s}\neq 0$ requires that $\gamma <1/2$, far
enough from $\gamma =1$ for noninteracting electrons. The
magnitude $|U_{s}|$ is determined by a competition between the
electronic gain of energy and its loss $\sim U_{s}^{2}$ from the
lattice deformation and charge redistribution. $3D$ ordering of
signs $U_{s}=\pm |U_{s}|$ discriminates the FE and anti-FE states.

\section{Conductivity, susceptibility and optics.\\
 Solitons, phasons and phonons.}

For a given $U_{s}$, the ground state is still doubly degenerate
between $\varphi =\alpha $ and $\varphi =\alpha +\pi $ allowing
for phase $\pi $ solitons which are the charge $e$ spinless
particles (holons) observed in conductivity at both $T\gtrless
T_{0}$. Also $U_{s}$ itself can change the sign between different
domains of ionic displacements. Then the electronic system must
also adjust its ground state from $\alpha $ to $-\alpha $ or to
$\pi -\alpha $. Hence the CD defect $U_{s}\Leftrightarrow -U_{s}$
requires for the phase soliton of the increment $\delta \varphi
=-2\alpha $ or $\pi -2\alpha $ which will concentrate the
noninteger charge $q=-2\alpha /\pi $ or $1-2\alpha /\pi $. Below
$T_{0}$, the $\alpha $- solitons must be aggregated \cite{teber}
into walls separating domains of opposite FE polarization; their
motion might be responsible for the observed frequency dispersion
of $\varepsilon $, which indeed is more pronounced bellow $T_{0}$
\cite{nad}. But at $T>T_{0}$ they may be seen as individual
particles which possibility requires for a fluctuational $1D$
regime of growing CD. It seems to take place sometimes as
demonstrated by the pronounced (while not singular in this case)
raise of $\varepsilon $ well above $T_{0}$ for the anti-FE case of
the $X=SCN$. It signifies the growing single chain polarizability
even before 3D interactions come to the game. But more typical
cases exclude the fluctuational regime: $\Delta $ increases
sharply below $T_{0}$ hence no pseudogap regime of the CD, also
the pure Curie-Weiss law in $\varepsilon $ extends widely around
$T_{0}$ signifying the mean field are over $\pm 30K$ .

Physics of soliton is particularly sensitive to a further symmetry
lowering and a very fortunate example is the subsequent AO of the
tetramerization in $(TMTTF)_{2}ReO_{4}$ \cite{anions,nad,bmn-02}.
The additional deformation exhorts upon electrons a $2k_{F}$ CDW
type effect thus adding the energy
\[
\sim {U_{ao}}\cos (\varphi -\beta )\cos \theta
\]
(here the shift $\beta $, mixing of site and bond distortions,
reflects the lack of the inversion symmetry below $T_{0}$). The
$U_{ao}$ term lifts the continuous $\theta -$ invariance thus
opening at $T<T_{ao}$ the spin gap $\Delta _{\sigma }\sim
U_{ao}^{2/3}$ as known for spin-Peierls transitions
\cite{spin-p,reidel-fukuyama}. Moreover it lifts even the discrete
invariance $\varphi \rightarrow \varphi +\pi $ of $H_{U}$ thus
prohibiting the $\pi $ solitons to exist alone; now their pairs
will be confined by spin strings. But the joint invariance
\[
\varphi \rightarrow \varphi +\pi \ ,\ \theta \rightarrow \theta +\pi
\]
is still present giving rise to \emph{combined topological
solitons} \cite{topdef} (cf. \cite{reidel-fukuyama}). Here they
are composed by the charge $e$ core (with $\delta \varphi =\pi $
within the length $\xi _{\rho }$ $\sim \hbar v_{F}/\Delta $) which
is supplemented by longer spin $1/2$ tails of the $\theta -$
soliton ($\delta \theta =\pi $ within the length $\xi _{\sigma
}\sim \hbar v_{F}/\Delta _{\sigma }\gg \xi _{\rho }$). These
complexes of two topologically bound solitons are the carriers
seen at $T<T_{ao}$ at the conductivity plot for the $X=ReO_{4}$,
see the plot in \cite{bmn-02}. Similar effects should take place
below intrinsically electronic transitions, particularly close can
be the spin-Peierls one for $X=PF_{6}$. But there the physics of
solitons will be shadowed by $3D$ electronic correlations which
are not present yet for the high $T_{ao}$ of $X=ReO_{4}$.

Contrary to a common interpretation, e.g.
\cite{bnl,optics,phonons}, the optical absorption edge is not a
two particle gap $E_{g}\neq 2\Delta$ but rather the collective
mode gaped at $\omega _{t}\approx \pi \gamma \Delta <2\Delta $.
(Here and below we simplify some relations as for $\gamma \ll 1$,
 see \cite{reidel-maki}.) The spectral region between $\omega _{t}$ and
 $2\Delta $ is filled by a sequence of quantum breathers, bound states of two
solitons. The regime changes qualitatively: from the essentially
quantum side $1/2<\gamma <1$ with $E_{g}=2\Delta$ to the quasi
classical low $\gamma $ scheme, just at the borderline for the CD
instability $\gamma =1/2$ which is not quite recognized in
existing interpretations of optical data. (E.g. the resent
extensive studies were all performed for the case equivalent to
$\gamma>1/2$, in our notations, and cannot be applied to the
$(TMTTF)_{2}X$ as it was supposed \cite{bnl}.) Notice also that
while the condition $\gamma <1/2$ follows ultimately from the
observed CD instability, the condition $\gamma <1/8$ for the
extreme suppression of charge fluctuations \cite{optics} is not
necessary, see more discussion below.

Since $\Delta$ is already well known and $\omega_{t}$ is
measurable, then we can access the basic microscopic parameter
$\gamma$. It is already clear that $\omega_{t}$ is much below
$2\Delta$, hence $\gamma$ is rather small, but the exact
determination of $\omega_{t}$ is complicated by phonon lines
present in the same region \cite{optics,phonons}. But these very
lines provide another, earlier unattended indication for the CD.
Their already noticed \cite{phonons} surprisingly high intensity
in $TMTTF$ may be due to the just lack of the inversion symmetry
lifted by the CD. (Oppositely, poorly resolved but still strong
sometimes, phonon lines in $TMTSF$ \cite{phonons} tell in favor of
a fluctuational regime of the CD in accordance with the pseudogap,
rather than a true gap, in electronic optical transitions.)

To respond to current needs of experimental analysis in optics we
shall present, without derivations, the formula for the mixed
electron-phonon contribution to the dielectric response function
valid at $T\geq T_{0}$:
\begin{equation}
\varepsilon (\omega )/\varepsilon _{\infty }=
1+\frac{(\omega_{p}^{\ast }/\omega _{t})^{2}(1-(\omega /\omega
_{0})^{2})}{(1-(\omega /\omega _{0})^{2})(1-(\omega /\omega
_{t})^{2})-Z}\;\;,\ Z=\left( \frac{\omega _{cr}}{\omega
_{t}}\right) ^{2-4\gamma }\leq 1
 \label{epsilon}
 \end{equation}
Here $\omega _{p}^{\ast }$ is the renormalized metallic plasma
frequency, $\omega _{0}$ is the bare frequency of a molecular
vibration associated to the CD, $\omega _{cr}(T)$ is the critical
value of the optical gap $\omega _{t}(T)$ below which the
spontaneous CD takes place. Near the criticality $Z(T_{0})=1$ we
see here the \emph{Fano antiresonance} at $\omega _{0}$, the
\emph{combined electron-phonon resonance} at
\[
\omega_{0t}^{2}\approx \omega_{0}^{2}+\omega_{t}^{2}
\]
and finally the \emph{FE soft mode} at
\[
\omega_{fe}^{2}\approx
\frac{(1-Z)}{\omega_{0}^{-2}+\omega_{t}^{-2}}.
\]
Being overdamped near $T_{0}$, this mode must grow in frequency at
$T<T_{0}$ following the order parameter, that is as $\sim
\varepsilon ^{-1/2}$, to become finally comparable with
$\omega_{0t}$. Near $T_{0}$ the two modes share the total weight
$\omega_{p}^{\ast 2}$ in the ratio $(\omega_{t}/\omega _{0})^{2}$
which is also the experimentally accessible parameter.

With reasonable suggestions on dependences $\omega _{t}(T)$ and
$\omega _{cr}(T)$ we find the critical singularity at $\omega =0$
as $\varepsilon (T)=A|T/T_{0}-1|^{-1}$. It develops upon the
already big gapful contribution $A\sim (\omega_{p}^{\ast }/\omega
_{t})^{2}\sim 10^{3}$ in a reasonable agreement with experimental
values $\epsilon \sim 10^{4}T_{0}/(T-T_{0})$. It confirms that the
FE polarization comes mainly from the electronic system, even if
the corresponding displacements of ions are very important to
choose and stabilize the long range $3D$ order.

The full quantitative implementation requires to resolve for
divergence (triple for the $TMTTF$ !) in reported values
\cite{optics,phonons} of such a basic, and usually robust,
parameter as the plasma frequency. The uncertainty could be simply
an artifact of inadequate parametrizations of
$\varepsilon(\omega)$ at the scale $\omega_{0t}$ (the right form
(\ref{epsilon}) was never exploited). But more fundamentally, it
can be also a signature of the strong renormalization
$\omega_{p}^{\ast}\ll\omega_{p}$ which could have developed while
the probe frequency decreases from the bare scale $\omega_{p}>1eV$
to the scale $\omega_{t},\omega_{0}\sim 10^{-2}eV$. Remind the
full (kinetic $\sim C_{kin}$ and potential $\sim C_{pot}$) energy
of elastic deformations (\ref{phases}) for the charge phase
$\varphi$:

\[
\frac{\hbar v_{F}}{4\pi }\left\{ (\partial _{t}\varphi
)^{2}C_{kin}/v_{F}^{2}+(\partial _{x}\varphi )^{2}C_{pot}\right\}
\equiv
\frac{1}{4\pi\gamma}\left\{(\partial_{t}\varphi)^{2}/v_{\rho}
+(\partial_{x}\varphi )^{2}v_{\rho }\right\}
\]
\[
\gamma =\frac{1}{C_{pot}C_{kin}}~,\ \frac{\omega _{p}^{\ast
}}{\omega _{p}} =\left( \gamma \frac{v_{\rho }}{v_{F}}\right)
^{1/2}=\frac{1}{C_{kin}}~,\ \frac{v_{\rho
}}{v_{F}}=\frac{C_{pot}}{C_{kin}}
\]
We see that the lowering of $\omega _{p}^{\ast }$ singles out the
effect of the effective mass enhancement $C_{kin}>1$ which is due
to coupling of the phase mode with acoustic phonons
\cite{fin-braz}. (Another factor for reduction of the parameter
$\gamma$, the Coulomb hardening $C_{pot}>1$ acts upon $\gamma$ and
 velocity $v_{\rho}$ \cite{harden} but cancels in their product
which gives $\omega_{p}^{\ast}$.) The mass enhancement will not be
effective above acoustic, or any other $q=0$, frequencies
$\omega_a$ (actually $C_{kin}=C_{kin}(0)$ is a function of
$\omega$:
$C_{kin}(\omega)=C_{kin}(0)\omega_a^2/(\omega_a^2+\omega^2$). It
explains the difference in extracting values of
$\omega_{p}^{\ast}$ from very high and from intermediate frequency
ranges.) If true, then the CD state resembles another Wigner
crystal: electrons on the $He$ surface, see \cite{shikin}, where
selftrapped electrons gain the effective mass from surface
deformations - the riplons.

\section{Perspectives and conclusions: the $TMTSF$ fate.}

Compounds of the $TMTSF$ subfamily are highly conductive which
today does not allow for low frequency experiments. Nevertheless
the transition may be their, just being hidden or existing in a
fluctuational regime like for stripes in High-$T_{c}$ cuprates
\cite{stripes}. (The fluctuational regime of the CD was already
observed in layered organic conductors \cite{Chiba-03}.) The
signature of the FE CD state may have been already seen in optical
experiments \cite{optics,dressel:96}. Indeed the Drude like peak
appearing within the pseudogap can be interpreted now as the
optically active mode of the FE polarization; the joint lattice
mass will naturally explain its, surprisingly otherwise, low
weight. Even the optical pseudogap itself
\cite{optics,dressel:96}, being unexpectedly big for $TMTSF$
compounds with their less pronounced dimerization of bonds, can be
largely due to the hidden spontaneous dimerization of sites.
Recall also the above mentioned optical activation of
intramolecular phonons \cite{phonons}.

A popular interpretation (see \cite{optics}) for optics of $TMTSF$
compounds neglects even the existent dimerization of bonds and
relies upon the generic 4- fold commensurability effects
originating higher order (8 particles) Umklapp processes. They
give rise to the energy $\sim U_{4}\cos 4\varphi$ which
stabilization would require for ultra strong $e-e$ repulsion
corresponding to $\gamma <1/8$ in compare to our moderate
constraint $\gamma<1/2$. While not excluded in principle, this
mechanism does not work in $TMTTF$, already because this scenario
does not invoke any CD instability. Moreover, the experiment shows
that even small increments of the dimerization, just below the
IInd order transition at $T_{0}$, immediately transfer to the
activation energy, hence the domination of the two-fold
commensurability.

Optical experiments will probably be elucidated when addressed to
members of the $(TMTTF)_2X$ family, showing the CD, with a
particularly reduced value of the associated gap (below typical
molecular vibrations - down to the scale of the pseudogap in
$(TMTSF)_2X$).

We conclude that the world of organic metals becomes polarized and
disproportionated. New events call for a revision (see more in
\cite{brazov-iscom}) of the existing picture (see \cite{bourbon})
and suggest new experimental and theoretical goals. Further
integrated studies are necessary. \bigskip

\textbf{Acknowledgments.} \newline
Author acknowledges collaboration with P. Monceau and F. Nad, discussions
with S. Brown, H. Fukuyama and H. Tajima; hospitality of the ISSP (Tokyo
University) and support from the INTAS grant 2212.

\end{document}